\documentclass{article}[15pt]      

\usepackage{amsmath,epsfig,color,calc,rotating,graphics}

\usepackage{natbib}
\bibpunct{(}{)}{;}{a}{}{,}

\usepackage[normalem]{ulem}

\begin{document}

\title{Quantifying Local Randomness in Human DNA and RNA Sequences Using Erd\"{o}s Motifs
\vspace{0.2in}
\author{Wentian Li$^1$, Dimitrios Thanos$^{2,3}$, Astero Provata$^3$\\
{\small  1. The Robert S. Boas Center for Genomics and Human Genetics}\\
{\small  The Feinstein Institute for Medical Research, Northwell Health, Manhasset, NY, USA}\\
{\small  2.  Department of Mathematics, National and Kapodistrian University of Athens, GR-15784 Athens, Greece}\\
{\small 3. Institute of Nanoscience and Nanotechnology, National Center for Scientific Research }\\
{\small  ``Demokritos", GR-15310 Athens, Greece}\\
}      
\date{}      
}  
\maketitle                   
\markboth{\sl Li et al. }{\sl Li et al. }

\vspace{-0.1in}

\large

\begin{center}
{\bf Abstract}
\end{center}

\normalsize
In 1932, Paul Erd\"{o}s asked whether a random walk constructed 
from a binary sequence can achieve the lowest possible deviation (lowest discrepancy), 
for the sequence itself and for all its subsequences formed by homogeneous arithmetic progressions.
Although avoiding low discrepancy is impossible for infinite sequences,
as recently proven by Terence Tao, attempts were made to construct such sequences with finite lengths. 
We recognize that such constructed sequences (we call these ``Erd\"{o}s sequences") exhibit certain hallmarks
of randomness at the local level: they show roughly equal frequencies of subsequences, and at the 
same time exclude the trivial periodic patterns. 
For the human DNA we examine the frequency of a set of Erd\"{o}s motifs of 
length-10 using three nucleotides-to-binary mappings. The particular length-10 
Erd\"{o}s sequence is derived by the length-11  Mathias sequence and is identical 
with the first 10 digits of the Thue-Morse sequence, underscoring the fact that both 
are deficient in periodicities. 
Our calculations indicate that: (1) the purine (A and G)/pyridimine (C and T) based Erd\"{o}s motifs are 
greatly underrepresented in the human genome, (2) the strong(G and C)/weak(A and T) based Erd\"{o}s 
motifs are slightly overrepresented, (3) the densities of the two are negatively correlated,
(4) the Erd\"{o}s motifs based on all three mappings being combined are slightly underrepresented, 
and (5) the strong/weak based Erd\"{o}s motifs are greatly overrepresented in the human messenger RNA sequences.

{\bf keywords:} Erd\"{o}s discrepancy problem; human genome; DNA motifs; local randomness;

\large

\newpage

\section{Introduction}      

\indent

The DNA sequences constituting the human genome are a product of long evolutionary history, with sequence
altering processes such as whole-genome duplication \citep{ohno68,wolfe},
regional (segmental) duplication \citep{bailey}, inversions \citep{flores},
insertion of foreign DNAs \citep{smit,cordaux}, insertion of mitochondrial DNA
to nuclear genome \citep{timmis,richly}, local insertions and deletions \citep{cooper,mills,payseur},
and most familiar to all, point mutations \citep{carlson}. 

Given our knowledge of these evolutionary processes, one might imagine a project to computationally simulate
the DNA sequence changes (e.g. \citep{wli-bc,koroteev}). However, there is a tremendous challenge
in choosing the correct model parameter values \citep{wli-jtb},
in modelling the natural selection \citep{carlos}, in putting the model in the
context of diploid with recombination \citep{duret}, and in a population \citep{hartl}. 
Point mutations lead to lesser destruction of the genome than larger-scale
changes, thus are more likely to survive to the next generation. Many point mutations
manifest as neutral mutations \citep{kimura}. If the point mutations dominate, the DNA sequences
would become more and more random.

There has been a debate on whether DNA sequences from the human genome should 
be considered to be random \citep{humgenome,clay2,iso}. 
If the DNA sequence from a complete chromosome is examined from one end to another, everybody agrees
it is not consistent with independent and identically distributed (iid) random variables
\citep{m1}. It was known that genome contains large domains with alternating high and low GC 
content called isochores \citep{bernardi85}.
Spectral analysis also confirmed that genome sequences are not white noise
\citep{voss,wli-mouse,wli-pre}.
For coding regions including the regulatory sites, codon structures as well as gene structure often
cause the DNA sequences to be non-random \citep{mani,yannis99,yannis02,hack,cocho}.
However, such debate has not really reached a consensus at the local level 
in non-coding, non-functional regions.  Part of
the reason is that randomness definitions are mostly based on probability models of infinitely long sequences, 
and corresponding definitions for finite sequences are not universally adopted.

Defining randomness in a sequence is not easy \citep{knuth}. On the one hand, for a sequence of infinite
length, being random implies ``any motif can appear". This includes the series composed of
the same symbol of arbitrary length, which by itself can not be considered as random.
On the other hand, for a finite sequence, the intuitive notion of being random
is that all symbols, dimers, triplets, etc. appear with equal frequency. However,
a periodic sequence would also satisfy this requirement, which nevertheless is hardly considered to be random.

In the field of mathematics, a discussion concerning the Erd\"{o}s 
discrepancy problem \citep{erdos9} seems to provide a foundation of randomness
in finite sequences.
Given an infinite binary sequence $\{ x_i \}$, taking values $x_i =(-1, 1)$, we 
define the discrepancy function for the sequence up to the length $L$, of an integer spacing $d$:
\begin{equation} 
\label{dis}
D(d,L) = \left| \sum_{i=1}^{k =  \lfloor L/d \rfloor}  x_{id} \right|.
\end{equation} 
In other words, $D(d,L)$ is the cumulative sum of the subsequence 
sampled from every $d$ positions, starting from the position $d$, of the length-L window.
Erd\"{o}s asked the question: 
given any constant $C$, can one find a sequence of length $L$ and a spacing value $d$ so that 
\begin{equation} 
D(d,L) > C 
\end{equation}
for one value of spacing $d$.
The problem is solved by Terence Tao,
and the answer is yes \citep{tao,sound}, no matter how large the value of C,
one can always find the above defined sum larger than C for certain spacing $d$ value, at a particular length $L$. 

When $d=1$, $ |\sum_{i=1}^{k=L} x_i |$ is just the cumulative function of a (random) walk
whose steps are given by the $\{ x_i \}$ series. In order for the Erd\"{o}s question
to have the negative answer, one should design steps to be as close to the origin as possible.
A periodic walk of period 2  with alternating positive and negative steps is the best solution.
However, this solution is unstable in the sense that as soon as one changes the $d$ value to 2,
$ |\sum_{i=1}^{k \le L/2} x_{2i} |$ diverges. Therefore, allowing $d$ to be different from
1, or, allowing one to sample other subsequences in $\{ x_{j=id} \}$ ($i=1,2,3, \cdots$) 
of spacing d (called ``homogeneous arithmetic progression"), is a solution to 
exclude periodic sequences from ``low discrepancy" (low cumulative value for the random walk)
category, while ``low discrepancy" is one  essential concept in our definition of randomness. 

We can reverse the Erd\"{o}s question to design a finite sequence which has the 
``lowest possible discrepancy" but exclude periodic sequence, for a given $C$ value. 
Such designed sequence exhibits the hallmark of a locally random sequence: uniform
distribution of single symbol and low-order tuples.  
It has been shown that there is a length-11 sequence which has the lowest possible
discrepancy, not larger than C=1 for any $d$ values \citep{mathias}. 
All other sequences with longer lengths would lead to discrepancy larger than 1. 
Similarly, for $C=2$, a length-1160 sequence is able to limit its discrepancy to equal 
or less than 2 for any $d$ \citep{konev}, but once the length is extended to 1161, 
discrepancy can increase above 2. We call these designed finite series ``Erd\"{o}s sequences with the
limit of C and of length L" ($E_{C,L}$). $E_{C,L}$ is our candidate of a locally random string at the
corresponding string length.

In this work, we will show that the length-11 sequence constructed by  Mathias
is not unique, whereas there exists a unique sequence with the required low-discrepancy feature at length 10.  
We focus on this length-10 sequence ($E_{1,10}$) and translate it to 10-mers in the DNA sequence
of the human genome. The length of 10 is particularly appealing as the DNA double helix makes
one complete turn after approximately 10 base-pairs (spacing between two ladder steps is 0.34 nm,
and one helix turn is 3.4 nm) \citep{calladine}. We will use the human genome
to examine the distribution of DNA substrings of length 10 which are consistent with the
$E_{1,10}$ sequence. The goal of this investigation is to examine whether the human genome tends
to possess these ``low discrepancy", ``locally random" sequences.

In the next section, we review the previously studied Mathias  $E_{1,11}$ sequence and propose 
the unique low discrepancy $E_{1,10}$ sequence. 
We translate the Erd\"{o}s $E_{1,10}$ motif to DNA sequences, using different base pair 
definitions. 
In section 3, the result section, we present the observed and expected $E_{1,10}$ values 
obtained from the human genome, and apply the same analysis to genomic sequences windowed 
by 1Mb windows to study its frequency at the local level. 
Then, the presence of $E_{1,10}$ sequences in particular functional DNA units and mRNA 
is discussed.  In the Discussion and conclusions section, the results are recapitulated 
and open problems are proposed, in particular the relation of $E_{1,10}$ and Thue-Morse sequences,
and the relevance of Kolmogorov complexity in defining locally random motifs.
In the Appendix comparison of the results in natural DNA with those of artificially constructed 
sequences are contrasted.

\section{Erd\"{o}s sequences}

\subsection{The unique length-10 Erd\"{o}s sequence with binary symbols}
 
\indent

\begin{figure}[th]
\begin{center}
     \epsfig{file=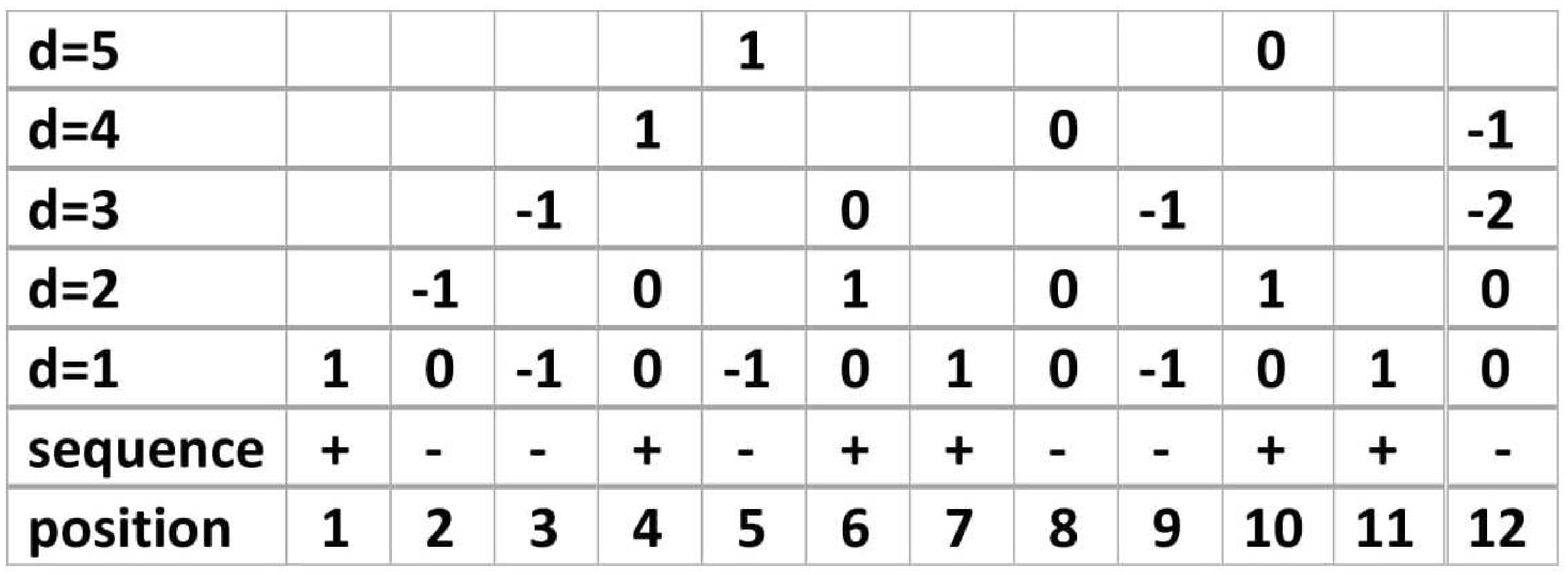, width=10cm}
\end{center}
\caption{
\label{fig1}
Illustration of $E_{1,11}$ sequence \citep{mathias}. The cumulative sum 
of $|\sum_{i=1}^{j < 12/d} x_{id}|$ at any $j$ (for d=1, 2, 3, 4, 5) is shown.
When the 12th value is $-1$, the cumulative sum is larger than 1 (in absolute value) for d=3.
If the 12th value is +1 (not shown), it would lead to the cumulative sum larger than 1 for d=1,2.
}
\end{figure}

The $E_{1,11}$ derived in  \citep{mathias} is {\sl + - - + - + + - - + + }, which is
called the ``Mathias sequence". 
Fig.\ref{fig1} illustrates why it satisfies Erd\"{o}s' low
discrepancy requirement: the sum of $|\sum_{i=1}^{i < 12/d} x_{id}|$ for d=1,2,3,4,5
is always bounded by 1. If a 12th value is added to the sequence, it can not be
+, as there would be three +'s in a row. If it is ``-", the deviation would be larger
than C=1 for d=3. Note that $|\sum_{i=1}^{i < 12/d} x_{1+(i-1)d}|$ has huge discrepancy
for $d=3$, reminding us that arithmetic progression has to start from position $d$.

Also note that if an Erd\"{o}s sequence reverses its direction, it may not be an
Erd\"{o}s sequence anymore. There could be two different causes. Take the
first 7 positions of $E_{1,11}$ for example (Fig.\ref{fig1}): in the forward direction, the
discrepancy series is (position 0 is also included): (0, 1,0, -1, 0, -1, 0, 1). 
In the backward direction, the discrepancy series is not a reverse of the above, 
but (0, 1,2,1,2,1,0,1). The two walks are mirror image of each other with respect
to 0.5. Only when the forward walk ends up at discrepancy of zero value, would
the two walks be mirror image of each other with respect to 0, and the backward
sequence to be also a Erd\"{o}s sequence.  The second cause is about ``phase". 
Since for $d > 1$, the subsequence selected starts from the position $i=d$,
forward and backward sequence may select different subsequences, thus potentially
different discrepancy series.

We can see that $E_{1,11}$ has a balanced single symbol frequencies (f(+)=6/11, f(-)=5/11),
and almost balanced dimer frequencies 
(\texttt{f(-{}-})=2/10, \texttt{f(-+)}=3/10,
\texttt{f(+-)}=3/10, \texttt{f(++)}=2/10). At the same time, $E_{1,11}$ is mostly not periodic.
Interestingly, when a sequence contains any k-mer exactly once (called De Bruijn
sequence \citep{bruijn,bruijn75}, its k-mer frequency is exactly balanced.  There are attempts to 
design shortest sequences to contain all k-mers in both direct and reverse direction of 
DNA sequences \citep{shortcake}. Again, to be the shortest means to contain any k-mer 
only once, therefore resulting  to a balanced k-mer frequencies.  
We do not expect Erd\"{o}s sequence to have a balanced $k$-mers frequencies
when $k$ is large, as the presence of the all-1 or all-0 $k$-mers  will lead to a large 
discrepancy for $d=1$. However, for $k$'s being less, equal, or close to $C$, we may
hypothesize balanced $k$-mer frequencies in Erd\"{o}s sequences. 

It is not difficult to check that there is another sequence
{\sl + - - + - + + - - + -}, denoted $E'_{1,11}$, with the last symbol changed from + to -,
which is also an Erd\"{o}s sequence.
It is because 11 is a prime number and  changing the last symbol only affects the discrepancy for $d=1$,
changing the final cumulative value from 1 to -1.  Swap + and - would not change any of the discrepancy
values. Running through all possible length-11 binary sequences, we observed that
Mathias' $E_{1,11}$ and $E'_{1,11}$ (plus the two derived from swapping the binary symbols)
are the only Erd\"{o}s sequences at $C=1$.

If the last symbol of the Mathias' sequence  can either be + or -, with the rest of the sequence identical,
we would expect that the first ten symbols should form a unique $E_{1,10}$. Indeed,
running through all possible length-10 binary sequences, we find only one sequence 
which conforms with the Erd\"{o}s discrepancy condition: {\sl + - - + - + + - - +}
(plus the one by swapping + and -). The single symbol
frequency is exactly balanced (f(+)=5/10, f(-)=5/10), whereas the dimer frequencies 
are somewhat unbalanced (\texttt{f(-{}-)}=2/9, \texttt{f(-+)}=3/9, \texttt{f(+-)}=3/9, 
\texttt{f(++)}=1/9).
Sequences with equiprobable subsequences are also called ``normal"
\citep{normal}.

Further simulations show that $E_{1, L}$ is unique when $L \le 10$ is even, and  not unique when
$L \le 11$ is odd, all with a degeneracy at the last symbol (can be either + or -). 
$E_{1, L}$ does not exist when $L > 11$. 
In the following, we focus on the unique Erd\"{o}s sequence with the maximum length, $E_{1,10}$,
and its frequency of appearance in the human genome.

\subsection{DNA motifs of length 10 bases which are associated with $E_{1,10}$ Erd\"{o}s sequence}

\indent

DNA sequences use four symbols (A,C,G,T). There are three different ways to split the
four symbols into two groups: 
(a) R/Y binarization combines AG (R for purine) and CT (Y for pyrimidine), 
(b) W/S binarization combines AT (W for weak) and CG (S for strong), 
(c) K/M binarization combines GT (K for keto) and  AC (M for amino).
The W/S binarization characterizes the binding strength between the two DNA strands
(G and C bind more strongly), and GC-content is an experimentally measurable quantity 
\citep{cscl} which is widely studied in genomic analysis \citep{gc}. The R/Y binarization
highlights the size difference of the two types of bases (A and G are larger in size),
and has been proposed to be relevant to codon patterns \citep{shepherd},
regulatory sequence patterns \citep{chris}, and the double helix structure \citep{arnott}.
The last binarization (M/K) is rarely used.
The fact that the DNA molecule has the form of a double helix implies that we must examine two strands
for Erd\"{o}s motifs, the direct strand, and the reverse complement strand.

(a) For R/Y binarization, four sequences are associated with $E_{1,10}$,
considering both the direct DNA sequence and its reverse complement
sequence on the opposite strand:
\begin{eqnarray}
\label{ry-based}
 1: & {\tt RYYRYRRYYR} \nonumber \\
 2: & {\tt YRRYRYYRRY} \nonumber\\
 3: & {\tt YRRYYRYRRY} \nonumber \\
 4: & {\tt RYYRRYRYYR} 
\end{eqnarray}
Seq.1 and seq.2 are two different mapping from +/- to R/Y; seq.3 is the reverse complement
of seq.1 (or reverse of seq.2); seq.4 is the reverse complement of seq.2 (or reverse of seq.1).
Note that seq.3 and seq.4 are not Erd\"{o}s sequences themselves, but their existence
indicates the presence of Erd\"{o}s sequence on the opposite strand of the DNA double helix.
As a reminder, the complementary rule is $A \leftrightarrow T$ and $G \leftrightarrow C$.

(b) Similarly, for W/S binarization, we have four motifs associated with $E_{1,10}$:
\begin{eqnarray}
\label{ws-based}
 1: & {\tt WSSWSWWSSW} \nonumber \\
 2: & {\tt SWWSWSSWWS} \nonumber \\
 3: & {\tt WSSWWSWSSW} \nonumber \\
 4: & {\tt SWWSSWSWWS}  
\end{eqnarray}
Seq.3 (seq.4) is the reverse complement of seq.1 (seq.2), noting that the complement operation 
maps one W (S) to another W(S).

(c) The least used  binarization is to M/K, and we again have four motifs:
\begin{eqnarray}
\label{km-based}
 1: & {\tt MKKMKMMKKM} \nonumber \\
 2: & {\tt KMMKMKKMMK} \nonumber \\
 3: & {\tt KMMKKMKMMK} \nonumber \\
 4: & {\tt MKKMMKMKKM}
\end{eqnarray}
In terms of the four nucleotide symbols, each motif represents $2^{10}=1024$
4-symbol patterns. Multiplying by  12, and 
subtracting 
48 = $16 \times 3$ 10-mers which belong to more than one type of motifs
(e.g. ACCACAACCA is either a RY or a WS motif), 
we are dealing with 
12240 10-mers which are associated with the Erd\"{o}s sequence $E_{1,10}$ 
in either one of the strands.

\section{Distribution of Erd\"{o}s sequences in the human genome}

\subsection{DNA sequence data}

\indent

In this study, the sequence of human reference genome  hg38 is used, which is downloaded from
UCSC Genome Browser {\sl http://hgdownload.soe.ucsc.edu/goldenPath/hg38/chromosomes/}.
We use chromosomes 1,2, $\cdots$ 22, and chromosome X but we exclude chromosome Y due to large
amount of non-sequenced regions.
The 23 sequences contain RepeatMasker filtering information: lowercase letters
represent regions that match transposons or other repetitive or low-complexity
sequences, whereas uppercase letters represent unique sequences.

The messenger RNA (mRNA)  sequence of ``known genes" (last updated in July 2016) 
is obtained from   
{\sl http://hgdownload.soe.ucsc.edu/goldenPath/hg38/database/}.
A mRNA sequence matches the genomic sequence of a gene, after removing
the introns. Due to alternative splicing, each genomic sequence at a gene locus
may contribute multiple mRNA sequences.
We obtained 197,783 mRNA sequences from the Genome Browser.

The refGene list of human genes (last updated in April 2018) is obtained from \\
{\sl http://hgdownload.cse.ucsc.edu/goldenPath/hg38/database/refGene.txt.gz}.
Pseudogenes are removed from the list by requiring the gene ending position
to be larger than the starting position.  Transcripts with overlapping coordinates
are merged using the bedtools program 
({\sl http://bedtools.readthedocs.io/}) \citep{bedtools}, with the command
{\tt bedtools merge -i file -d 100 -c 4 -o collapse}. 
This processing leads to 18,757 gene sequences.

The non-transposon/non-repetitive unique sequence can be read out from the 
reference genome directly: unique sequences are in uppercase, whereas RepeatMasker \\
({\sl http://www.repeatmasker.org/}) identified repetitive sequences are in lowercase.

The telomere and centromere region is first based on the
cytoband information from \\
{\sl http://hgdownload.soe.ucsc.edu/goldenPath/hg38/database/cytoBand.txt.gz}.
The first and the last bands of a chromosome is considered to
be the telomere regions. We first use the same file to bracket the centromere
region (when the band is labeled as ``acen"). Then we further fine-tune the
boundary by an observation made in \citep{thanos} that windowed statistical
qualities (e.g. entropy) have extremely low variations in the centromere region.

\subsection{Distribution of Erd\"{o}s sequences in the human genome at the chromosome level}

\indent

For R/Y binarization, out of 2.911 billions overlapping 10-mers in the human genome (chromosomes 1-22,X,
excluding any 10-mers which contain unsequenced bases) 
there are 6,161,338 counts of R/Y $E_{1,10}$, or 0.21\% of all 10-mer counts. 
The ratio between R/Y $E_{1,10}$ and non-R/Y-$E_{1,10}$ sequence counts is 1:471.
This frequency is severely underrepresented as the expected frequency is 0.39\%
if the strand symmetry holds true (see Appendix, from both analytic formula
and simulation results). The observed over expected
ratio (O/E) is 0.54, or equivalently 1/1.85. 
On individual chromosomes, the R/Y observed frequency of $E_{1,10}$ compatible
10-mers is also much lower than those in sequences 
generated randomly using the observed base composition (see Appendix).

For W/S $E_{1,10}$, there are 
10,073,985 copies of them in the human genome (0.35\%). For each W/S $E_{1,10}$, there are 
288 10-mers that are not $E_{1,10}$ compatible.  This frequency is slightly higher than the expected 
value of 0.32\% (O/E=1.05), if we make a simple assumption that the GC-content is 40\% (see Appendix). 
The 40\% value is a good approximation for global GC-content \citep{gc}.
Even if we use the observed GC-content at the individual chromosome, the conclusion
remains true that the frequency is slightly higher than expected (except for chromosomes
16 and 19) (see Appendix).

Finally, there are 8,782,253 copies of M/K based $E_{1,10}$ (0.3\%), slightly lower than
expected (0.39\%, assuming strand symmetry).
There are more W/S  or M/K $E_{1,10}$ than R/Y $E_{1,10}$,
even though R/Y $E_{1,10}$ is expected to appear more often than W/S $E_{1,10}$,
and equally likely as M/K $E_{1,10}$.
Combining the three types of $E_{1,10}$, there are in total 
24,692,591 copies of 10-mer motifs associated with $E_{1,10}$.  This number is slightly lower than the sum of 
the 
three counts for R/Y, W/S, and K/M based $E_{1,10}$ because
some 10-mers belong to more than one type. The frequency for overall $E_{1,10}$ is 
0.85\%. This frequency is lower than the expected value of 1.1\% (see Appendix).

There is yet another R/Y based 10-mer \citep{trifonov10,wli-motif},
RRRRRYYYYY /YYYYYRRRRR (R5Y5), proposed as a nucleosome positioning sequence or motif
\citep{trifonov80,drew,peckham,widom,jiang}.
There are other nucleosome positioning sequence patterns proposed, in particular,
the periodicity 10-11 of AA/TT steps \citep{calladine}.
Unlike the underrepresented R/Y-based Erd\"{o}s sequences, 
the R5Y5 motif is overrepresented: observed frequency is 0.36\% vs. the expected 0.195\%
(O/E=1.87).
The overrepresentation of R5Y5 is also consistent with the abundance of R- and Y-tracts \citep{behe},
and their connection to coding/noncoding regions has been reported \citep{yannis97}.

Table 1(a) shows the correlation among various $E_{1,10}$'s. The R/Y based Erd\"{o}s frequency
is negatively correlated with both W/S based and K/M based Erd\"{o}s frequencies, at the chromosome level,
with Spearman rank correlation coefficient (\citep{hollander}, sec. 8.6)
of $-0.88$ and $-0.64$. The W/S based and K/M based Erd\"{o}s
sequences are positively correlated. The overall Erd\"{o}s frequency is closely tied
to that of the W/S based Erd\"{o}s sequence,
probably because there are more W/S based Erd\"{o}s motifs than either 
R/Y based or K/M based ones.

Table 1(b) shows the correlation between Erd\"{o}s sequence frequencies with
those of other sequence features. The R/Y based Erd\"{o}s sequence is always opposite
to W/S based or to K/M based, as well as overall Erd\"{o}s sequence, in terms of its correlation
with other features. Notably, the rate of W/S based Erd\"{o}s sequence is positively 
correlated with the GC-content, and the frequency of R/Y based Erd\"{o}s sequence is positively
correlated with the chromosome length.

For completeness, we also show the correlation among the frequencies of various
sequence features in Table 1(c). Interestingly (and perhaps counter-intuitively), 
both the polyA/polyT density and polyC/polyC density (which is very low) are 
positively correlated with the GC-content.

\begin{figure}[th]
\begin{center}
     \epsfig{file=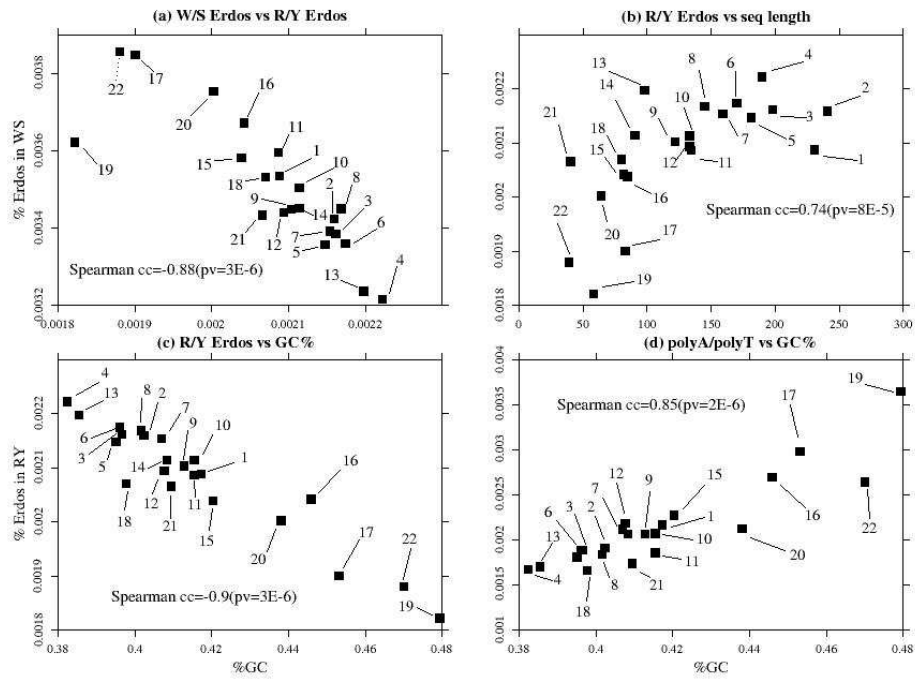, width=16cm}
\end{center}
\caption{
\label{fig2}
Each point represents the average of a chromosome.
(a) x: frequency of R/Y Erd\"{o}s sequence $E_{1,10}$,
y: frequency of W/S Erd\"{o}s sequence.
The chromosome number is marked. The grey line indicates the equality between x and y. 
(b) x: sequence length (excluding unsequenced bases), y: R/Y based Erd\"{o}s $E_{1,10}$ frequency.
(c) x: GC-content, y: frequency of R/Y based Erd\"{o}s sequence $E_{1,10}$.
(d) x: GC-content, y: frequency of polyA/polyT 10-mers. 
}
\end{figure}

Fig.\ref{fig2} highlights a few strong correlations: negative correlation between
two types of Erd\"{o}s $E_{1,10}$'s,  positive correlation between R/Y based $E_{1,10}$
and sequence length,  negative correlation between R/Y based $E_{1,10}$ and GC-content,
and positive correlation between polyA/polyT density and GC-content. 
Other large (in absolute value) and significant correlations are marked as bold in Table 1.

\subsection{Distribution of Erd\"{o}s sequences in the human genome at the window level}

\indent

Quantities calculated at the chromosome level may not be fine scaled enough. In order to check
whether Erd\"{o}s sequence frequencies are correlated with other sequence features at a more
local scale, we partition the human genome into non-overlapping 1Mb windows.
Similar to Table 1, we show Spearman correlation in three categories, i.e, 
among various Erd\"{o}s sequence frequencies (R/Y, W/S, K/M based, and overall $E_{1,10}$), 
between Erd\"{o}s sequence and other sequence features, and among sequence features themselves
(GC-content, Y5R5 motif frequency, non-transposon/non-repetitive sequence frequency, 
length-10-polyA/polyT frequency, and length-10-polyC/polyG frequency). 

\begin{figure}[th]
\begin{center}
  \begin{turn}{-90}
     \epsfig{file=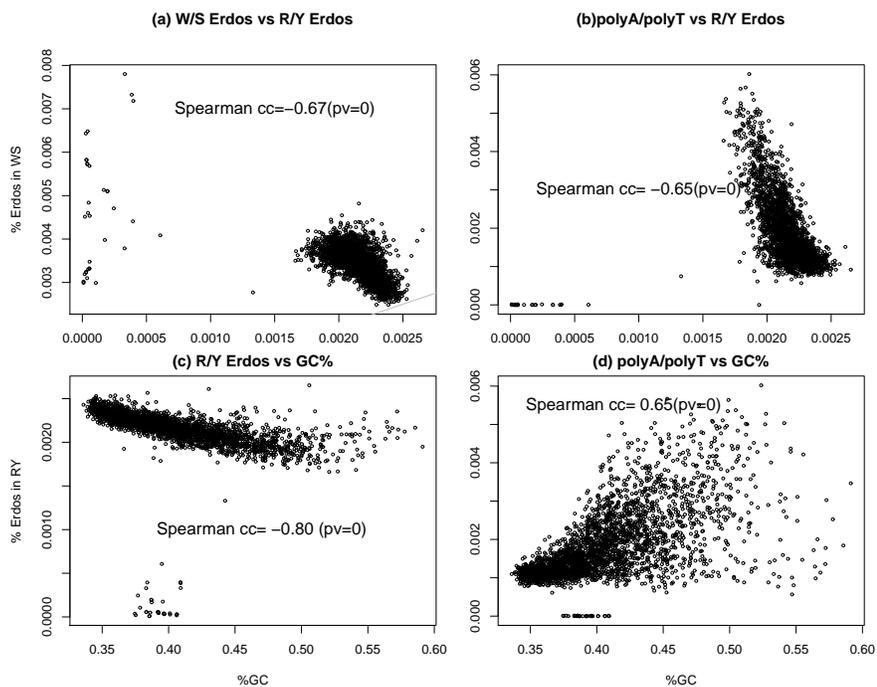, width=9cm}
  \end{turn}
\end{center}
\caption{
\label{fig3}
Each point represents a 1Mb window (a,c,d can be compared with those in Fig.\ref{fig2}).
(a) x: frequency of R/Y based Erd\"{o}s sequence $E_{1,10}$,
y: frequency of W/S Erd\"{o}s sequence.
The grey line indicates the equality between x and y. 
(b) x: frequency of R/Y based Erd\"{o}s sequence $E_{1,10}$,
y: frequency of polyA/polyT 10-mers.
(c) x: GC-content, y: frequency of R/Y based Erd\"{o}s sequence $E_{1,10}$.
(d) x: GC-content, y: frequency of polyA/polyT 10-mers. 
}
\end{figure}

Interestingly, all highlighted strong correlations in Table 1 and Fig.\ref{fig2} at the chromosome level
remain true at the 1Mb window level (Table 2, Fig.\ref{fig3}), including the negative correlation 
between R/Y based and W/S (and K/M based) Erd\"{o}s sequence frequency, the negative (positive) 
correlation between R/Y (W/S and overall) Erd\"{o}s sequence frequency and GC-content,
the positive correlation between polyA/polyT and GC-content, etc.
The scatter plots in Fig.\ref{fig3} also show the existence of outliers which are all from centromere regions. 
It also shows that the positive correlation between polyA/polyT and GC-content coexist with 
a larger variance for polyA/polyT at GC-rich regions (see Fig.\ref{fig3}(d)).
The statistical test results in Table 2 are all more significant than the corresponding
ones in Table 1, as the sample size is increased from the 23 chromosomes to 2755 1Mb windows.

The overrepresentation of W/S based $E_{1,10}$ sequence observed at chromosome level is still true
at 1Mb window level. We use the local GC-content to estimate the expected frequency of
W/S based $E_{1,10}$ in each window. The distribution of all O/E is single-peaked, with 
median 1.097, mean 1.091, mode around 1.12-1.13. Of all 1Mb windows, 86\% overrepresent
W/S $E_{1,10}$ and 14\% underrepresent it. As a comparison, none of the 1Mb window 
overrepresents R/Y based $E_{1,10}$, with mean of O/E to be 1/1.827 (if outliers removed,
1/1.81), median 1/1.796 (if outliers removed 1/1.794).

In \citep{wli-motif} it was observed that the R5Y5 density is negatively correlated with 
that of the transposon and other repetitive sequences at 64kb window level. Table 2 shows
that this conclusion remains true at the 1Mb window level (positive correlation between R5Y5
density and unique sequence density).

\subsection{Frequency of Erd\"{o}s motifs in various functional, regional, sequence-feature classes}

\indent

Besides counting Erd\"{o}s motifs in the whole genome, we also count them in  
sequences belonging to specially defined sub-categories.
The first subcategory consists of the genome free of transposons and repeats.  
This sequence has been formed by applying the RepeatMasker tool over the genome and 
is represented as uppercase letters in the reference genome.
Messenger RNA (mRNA) sequence are concatenated exons from the same gene.
The gene sequences are the genomic sequences
bracketed by the starting and ending position of genes as listed in refGene,
which contains both exons and introns. 
Telomeres and centromeres exhibit very different sequence features
compared to the rest of the sequences, and can also be examined separately. 

\begin{figure}[th]
\begin{center}
  \begin{turn}{-90}
     \epsfig{file=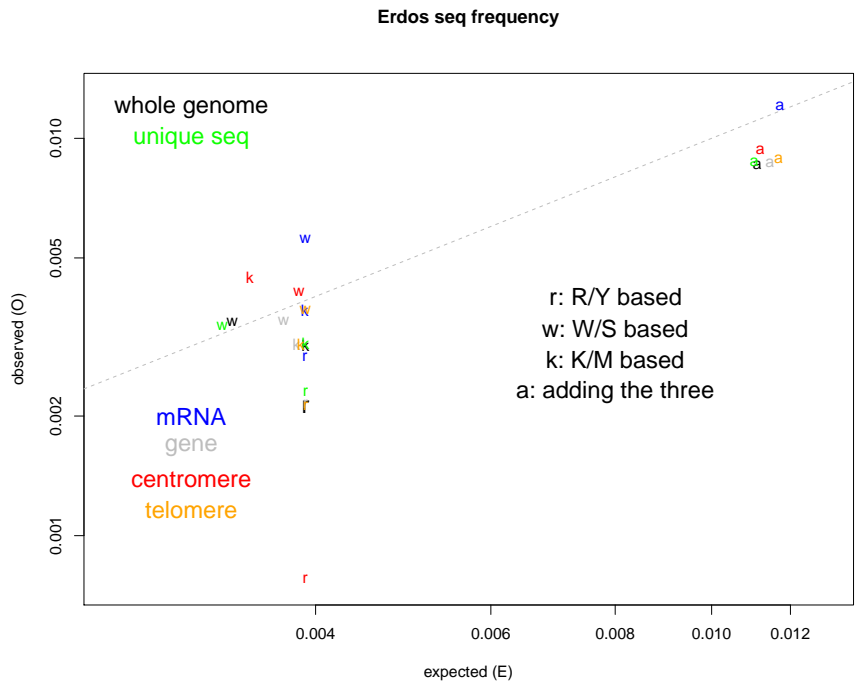, width=9cm}
  \end{turn}
\end{center}
\caption{
\label{fig4}
Comparison of the observed Erd\"{o}s sequence frequency (y-axis) and the expected (x-axis).
The letter labeling the dots indicates the type: r for R/Y based, w for W/S based, k for K/M based
$E_{1,10}$, ``a" is the sum of the three. The color indicates the source of the sequences: whole genome,
non-transposon-non-repetitive unique sequences (green), mRNA sequences (blue), gene sequences (grey),
centromere (red) and telomere (yellow) sequences. The expected frequency is based on the product
of single symbol frequencies obtained from the examined sequence.
}
\end{figure}

Fig.\ref{fig4} summarizes the observed vs expected Erd\"{o}s motifs in the following situations:
(1) non-transposon, non-repetitive ``unique" sequences; (2) messenger RNA sequences; 
(3) gene sequences; (4) centromeres; and (5) telomeres. 
The expected frequencies derived by the formulae in Appendix use 
the exact S\%, R\%, K\% frequencies as obtained from the same categories, 
instead of the equiprobability assumption for R/Y and K/M based $E_{1,10}$
sequence, and GC\%=0.4 assumption for W/S based one.

Fig.\ref{fig4} shows that the underrepresentation of R/Y-based Erd\"{o}s motifs
is true for all categories examined, in particular the centromere region. However,
the extreme lower R/Y based Erd\"{o}s motif frequency in centromere might be
an artifact. Centromeres are dominated by a large number of copies of the alpha satellite
sequence of length 172 bases. Using the expected frequencies, the expected number of
R/Y based Erd\"{o}s motif is 0.63 out of 162 10-mers. The actual number of observed copies is zero. 
As a comparison, the expected number of W/S based Erd\"{o}s motif is 0.52, but there is
only one observed copy. Since both zero and one copy are consistent with the expectation, R/Y based
motif just happens to be on the lower side of the expectation.

For W/S based Erd\"{o}s motifs, most observed frequencies are consistent or slightly
higher than the expected (Fig.\ref{fig4}), except for mRNA sequences. We hypothesize
that it might be related to a ``hidden" periodicity in the $E_{1,10}$ sequence.
If we examine the $E_{1,10}$ sequence closely in Fig.\ref{fig1}, the positions 1, 4, 7, 10 are
all positive, resulting in a periodicity of three in this particular reading frame.
Similarly, the positions 2, 5, 8 are all negative, again a potential periodicity of three. 
The reason that this local periodicity evades the attention in the mathematics community 
in the discussion of Erd\"{o}s problem is that the homogeneous arithmetic progression requires the
step-$d$ walk to start from position $d$. When $d=3$, this requirement forces
the walk to start from the third reading frame, not the first or the second reading 
(see Fig.\ref{fig1}). Since mRNA sequences contain periodicity-three signals, 
it is not unreasonable to speculate that it is a possible
cause of the Erd\"{o}s sequence enrichment. In addition,  mRNAs contain specific 
motifs of codons; these may also be mirrored in the Erd\"{o}s sequences and thus 
contribute further in the  observed enrichment of W/S-Erd\"{o}s motifs.

Finally, adding the three types of Erd\"{o}s frequency together, for both expected
and observed sequences, we see in Fig.\ref{fig4} a pattern of underrepresentation (with
the exception of mRNA). This underrepresentation, generally speaking, supports the
idea that the human DNA sequences are not locally random.

\section{Discussion and conclusions}

\indent

There is a long history in characterizing statistical patterns in DNA sequences:
from recognizing the nearest neighbor correlations \citep{kornberg} to the
detection of long-range correlations \citep{likaneko,peng,voss}. The debate on whether
the human genome sequences are homogeneous, iid (independent and identically distributed),
random \citep{humgenome,clay2,iso} is often based on concepts defined on
infinitely long sequences. There is always a lack of adequate concepts related
to randomness on finite scales. In this study, our aim is to consider a type of 
low-discrepancy sequences, called Erd\"{o}s sequences $E_{C,L}$, 
and perhaps their generalization, as candidates for locally random sequences.

The underrepresentation of overall $E_{1,10}$ compatible motifs in the human genome
(Fig.\ref{fig4}) can be interpreted as a lack of locally random sequences.
This is particularly true for R/Y based $E_{1,10}$ 10-mers. Interestingly,
it was observed that long-range correlation scaling is best observed in 
the R/Y binarization \citep{peng}. The surprising overrepresentation
of W/S based $E_{1,10}$ 10-mers in mRNA sequences might reflect, 
among others, a potentially
periodicity-3 tendency in certain reading frames in this sequence. Further
investigation is required to pinpoint the cause of the overrepresentation.

The underrepresentation of R/Y based $E_{1,10}$ is not a consequence of
lower or higher R content, or equivalently, a violation of  Chargaff's second
parity. Indeed, if R\% is too high or too low, other R/Y-based 10-mers
may become the most frequent motifs, such as R-track or Y-track
\citep{yannis97}. However,  Chargaff's rule is well preserved both at the
chromosomal and at the 1Mb window levels (result not shown). Furthermore, we
observed that among R/Y based 10-mers with R\%=Y\%, the Erd\"{o}s motifs
are still underrepresented (result not shown).

The key component in an Erd\"{o}s sequence is its low discrepancy, either in
the direct cumulative plot or in  many of its equally spaced subsequences.
The cumulative plot or (random) walk representation has been frequently used
in DNA sequences \citep{skolnick,peng,zhang}, but the inclusion of equally spaced
subsequences is an ingenious device to exclude periodic sequences.
An open question is how low discrepancy property fits other hallmarks of local randomness.

One feature of finite random sequences is their lack of periodic patterns.
Indeed, $E_{1,10}$ is identical to the first 10 digits of the Thue-Morse sequence,
a well known aperiodic or quasiperiodic infinite sequence
\citep{allouche-TM,riklund}. Interestingly, Thue-Morse sequence
has bounded discrepancy for specific spacing values ($d=1,2,4,8, \cdots$) 
even in the infinite sequence length limit \citep{leong-master,leong}.

Using a semi-periodic pattern to reduce discrepancy while at the same time
to avoid exact periodicity may have other applications in sequence analysis. 
For example, it is well known that in the promoter region of housekeeping genes,
CpG dinucleotide is common, forming CpG islands \citep{vinson}.
However, they can not be arranged in a periodic fashion. When they appear in
a periodic arrangement of CGG repeats with more than 200 copies in gene FMR1,
it leads to a form of intellectual disability \citep{park}.

Another property of local randomness is the difficulty in recreating the
sequence. This property makes randomness equivalent to a measure of complexity
\citep{wli-comp}. The use of Kolmogorov complexity as a way to measure local
randomness is an involved topic \citep{allouche,vitanyi,kol-book,soler} and will not be addressed here.
However, compressibility based calculation can be easily carried out
\citep{zl77,benedetto,estevez}. Preliminary calculations indicate that $E_{1,10}$ belong to a 
harder-to-compress group, but not the hardest-to-compress (results not shown).

In conclusion, the search for low discrepancy sequences in the human DNA has 
shown that these motifs are overall underrepresented (except for some particular cases). 
This results is consistent with nonrandomness at the local scale. 
In future studies, it would be interesting to extend this search into the full 4-letter alphabet scheme, 
Erd\"{o}s sequences of longer sizes, both of which require considerably more involved 
computational effort, as well as extending low discrepancy property to all reading frames. 
Knowing that DNA sequences are not homogeneous, but are composed by
sub-regions (isochores, genes, etc.), it would be interesting to quest 
whether these inhomogeneities are also mirrored locally following the
randomness criteria as posed by the Erd\"{o}s problem.

{\bf Acknowledgement:} WL thanks the Robert S Boas Center for Genomics and Human Genetics
for support.

\section*{Appendix: estimation of the $E_{1,10}$ frequencies in random sequences}
\indent

If each symbol appears with equal probability, the expected $E_{1,10}$ frequency can
be estimated by counting the number of 4-symbol sequences matching $E_{1,10}$
of the total number of possible 10-mers. There are 4 motifs in Eq.(\ref{ry-based}), each containing
$2^{10}=1024$ 4-symbol sequences. Combining Eqs.(\ref{ry-based},\ref{ws-based},\ref{km-based}), there are roughly 
$1024 \times 3 \times 4$ =12288 4-symbol 10-mers associated with $E_{1,10}$ 
(the actual number is 12240 because one 4-symbol sequence can simultaneously 
belong to (e.g.) Eq.(\ref{ry-based}) or Eq.(\ref{ws-based}) ). The total number of 4-symbol 10-mers is $4^{10}=1048576$.
Therefore the $E_{1,10}$ associated 10-mer frequency is 12240/1048576 $\approx 1.17\%$.

For individual type of $E_{1,10}$ sequence, R/Y based Erd\"{o}s sequence
is expected to appear with the frequency of 
$$ E[P_{R/Y}]= 4 \times R\%^5 Y\%^5.$$
If the strand symmetry \citep{fickett,li97,forsdyke} holds true,
i.e., A\% $\approx$ T\%, G\% $\approx$ C\%, we have R\% $\approx$ Y\% $ \approx 0.5$.  
Then the expected frequency is $4 \times (1/2)^{10}= 0.39\%$.
The same argument is also applied to K/M based Erd\"{o}s motifs.
For W/S based Erd\"{o}s sequence, the expected frequency is 
$$ E[P_{W/S}]= 4 \times W\%^5 S\%^5. $$
If the strand symmetry is true, S\% $\approx$  2 G\% $\approx$ 2C\%,
W\% $\approx$  2 A\% $\approx$ 2T\%.
In the  human genome, it is observed that G\% $\approx$ C\% $\approx$ 0.2,
and A\% $\approx$ T\% $\approx$ 0.3 \citep{gc}. The expected W/S 
based $E_{1,10}$ frequency is $4 \times 0.4^5 \times 0.6^5 = 0.3185\%$.

We also ran a simulation to generate artificial chromosomes with the
same base composition as the real human chromosomes, but bases are
scrambled. The numbers of R/Y and W/S based Erd\"{o}s sequence $E_{1,10}$
in the simulated artificial chromosomes are shown in Table A1 
(column ``exp(simu)". These can be compared to the estimated from the formula
(``exp(form)"). These two columns match very well. In comparison,
the observed numbers of R/Y based Erd\"{o}s sequences in the human genome
are greatly underrepresented, whereas W/S based Erd\"{o}s sequences are
slightly overrepresented.

\begin{center}
\begin{tabular}{c|ccc|ccc}
\hline 
ch & \multicolumn{3}{c|}{R/Y based} & \multicolumn{3}{c}{ W/S based} \\ 
 & exp(form) & exp(simu) & obs & exp(form) & exp(simu) & obs\\
\hline 
1&900310  & 897577&481226&783565  & 784613&814642 \\
2&939640  & 939599&519480&773397  & 773140&823550 \\
3&773828  & 773843&428228&622072  & 620791&670565 \\
4&741221  & 740585&421683&557761  & 557947&610107 \\
5&708065  & 709136&389259&565271  & 566846&608698 \\
6&664369  & 663453&369791&532683  & 533044&571511 \\
7&620975  & 620758&342407&520689  & 518623&539313 \\
8&565499  & 564884&313915&464086  & 463694&499486 \\
9&475743  & 475759&256150&407606  & 408054&420120 \\
10&520555  & 519739&281705&450212  & 449989&466925 \\
11&525522  & 526585&280736&454480  & 454826&483844 \\
12&520068  & 520663&278794&437233  & 436291&458118 \\
13&382746  & 381916&215344&292380  & 292947&317177 \\
14&353779  & 353633&191478&298201  & 298697&312635 \\
15&330629  & 330103&172558&290753  & 290061&303179 \\
16&319552  & 319783&167038&301244  & 301711&300467 \\
17&323904  & 324392&157592&309935  & 309961&319059 \\
18&312844  & 311656&165815&252685  & 252467&282838 \\
19&228284  & 228618&106470&226350  & 225640&211684 \\
20&249779  & 250204&128054&231185  & 231051&240010 \\
21&156594  & 156317  & 82809&132512  & 132147&137638 \\
22&152961  & 153855  & 73622&150241  & 149192&151017 \\
X&605050  & 604301&337184&483427  & 482816&531402 \\
\hline 
\end{tabular}
\end{center}
Table A1: The number of R/Y and W/S-based Erd\"{o}s sequences in 1-22 \& X chromosomes. 
The counts are based on the formulae in Appendix (left columns), 
artificially constructed chromosomes (middle columns), and the true observed 
numbers calculated from the human genome (right columns).

\newpage

\newpage
\begin{table}[ht]
\begin{center}
(a) correlation among various Erd\"{o}s sequence frequencies \\
\begin{tabular}{c|ccc}
\hline
 & \% W/S Erd\"{o}s & \% K/M Erd\"{o}s & overall \% Erd\"{o}s \\
\hline
\% R/Y Erd\"{o}s & {\bf -0.88 (2.8E-6)} & -0.64 (0.0014) & {\bf -0.75 (5.5E-5)} \\ 
\% W/S Erd\"{o}s & & 0.60 (0.0028) & {\bf 0.89 (3.2E-6)} \\ 
\% K/M Erd\"{o}s & & &		{\bf 0.78 (2.2E-5)} \\ 
\hline
\end{tabular}
\\ [0.4in]
(b) correlation between Erd\"{o}s sequence and other sequence features \\
\begin{tabular}{c|cccccc}
\hline
 & GC\% &  uniq-seq\% & L0-polyA/T \% & L10-polyC/G \%  & R5Y5\% &  L(sequenced)\\
\hline
R/Y Erd\"{o}s & {\bf -0.90 (3E-6)} & 0.5 (0.02)& -0.69 (4E-4) & -0.71(2E-4) &-0.37(0.08) & {\bf 0.74 (8E-5)} \\
W/S Erd\"{o}s & {\bf 0.91 (3E-6)} & -0.52(0.01)& 0.68(5E-4)& 0.71(2E-4)& 0.38(0.08) & -0.65 (0.001)\\
K/M Erd\"{o}s & 0.59(0.004) & -0.33(0.1) & 0.52(0.01) & 0.38(0.08) & 0.19(0.37)& -0.24(0.27) \\
overall Erd\"{o}s & {\bf 0.78(2E-5)} & -0.35(0.1)& 0.62(0.002)& 0.53(0.01)& 0.36(0.09) & -0.40 (0.06)\\
\hline
\end{tabular}
\\ [0.4in]
(c) correlation among sequence features \\
\begin{tabular}{c|ccccc}
\hline
 & uniq-seq\% & L0-polyA/T \% & L10-polyC/G \%  & R5Y5\% &  L(sequenced)\\
\hline
GC \% & -0.49(0.02) & {\bf 0.85(2E-6)} & {\bf 0.87(3E-6)} & 0.21(0.3)& -0.61(0.003)\\
 uniq-seq \% & & -0.40(0.06)& -0.48(0.02)& -0.21 (0.4)& 0.37(0.08)\\
L0-polyA/T \% & & & {\bf 0.78(2E-5)} & -0.092(0.7)& -0.35(0.1)\\
L0-polyC/G \% & & & & 0.16(0.5)& -0.54(0.008)\\
R5Y5 \% & & & & & -0.33(0.1)\\
\hline
\end{tabular}
\end{center}
\caption{
(a) Spearman correlation (and the corresponding p-value for testing its value being zero) among various
Erd\"{o}s $E_{1,10}$ sequence frequencies: R/Y based, W/S based, K/M based, and overall Erd\"{o}s sequences. 
(b) Spearman correlation (and the p-values) between Erd\"{o}s $E_{1,10}$ frequencies
and other chromosome sequence features: GC-content,  non-transposon, non-repetitive ``unique"
sequence frequency,  frequency of poly-A or poly-T of length 10, frequency of poly-C or poly-G of length 10,
a R/Y based nucleosome positioning motif, chromosome length (excluding the unsequenced bases).
(c) Spearman correlation (and the p-values) among various sequence features at the chromosome level.
Test results are are highlighted in bold if the $p$-value is lower than 0.001, following
a suggestion by \citep{colquhoun}.
}
\end{table}

\newpage
\begin{table}[ht]
\begin{center}
(a) correlation among various Erd\"{o}s sequence frequencies \\
\begin{tabular}{c|ccc}
\hline
 & \% W/S Erd\"{o}s & \% K/M Erd\"{o}s & overall \% Erd\"{o}s \\
\hline
\% R/Y Erd\"{o}s & -0.68(0) & -0.58 (3.6E-249) & -0.53 (6.3E-199) \\
\% W/S Erd\"{o}s & & 0.43 (4.2E-123) & 0.93 (0) \\
\% K/M Erd\"{o}s & & & 0.56 (7.8E-231) \\
\hline
\end{tabular}
\\ [0.4in]
(b) correlation between Erd\"{o}s sequence and other sequence features \\
\begin{tabular}{c|ccccc}
\hline
 & GC\% &  uniq-seq\% & L0-polyA/T \% & L10-polyC/G \%  & R5Y5\%   \\
\hline
R/Y Erd\"{o}s & -0.80 (0) & 0.029 (0.13)     & -0.65(0)  & -0.29(1.5E-52) &  -0.15 (2.4E-15)\\
W/S Erd\"{o}s & 0.87 (0)  & 0.089 (3.2E-6)   & 0.37 (4.3E-92) & 0.24 (2.5E-36) &  0.51 (3E-181)\\
K/M Erd\"{o}s & 0.56(7.5E-229) & 0.18 (2.3E-22)& 0.50 (2.1E-173) & 0.22 (4.2E-31)&  -0.0052 (0.79)\\
overall Erd\"{o}s & 0.80 (0) & 0.17(2.6E-19) & 0.33 (8.5E-70) & 0.21 (2E-29) &  0.49 (6.5E-165) \\
\hline
\end{tabular}
\\ [0.4in]
(c) correlation among sequence features \\
\begin{tabular}{c|cccc}
\hline
 & uniq-seq\% & L0-polyA/T \% & L10-polyC/G \%  & R5Y5\% \\
\hline
GC \% & 0.090 (2.1E-6)  & 0.65(0) & 0.37 (2.4E-92) & 0.28 (3.2E-52) \\
uniq-seq \% &  & -0.091 (1.9E-6) & -0.0043 (0.82) & 0.30 (8.2E-60) \\
L0-polyA/T \% &  & & 0.35(1.6E-80)  & -0.25(2.9E-41) \\
L0-polyC/G \% &  & & & -0.038 (0.047) \\
\hline
\end{tabular}
\end{center}
\caption{
Similar to Table 1 for variables calculated at the non-overlapping 1Mb window.
Note that chromosome length in Table 1 is no longer a variable here.
}
\end{table}

\end{document}